\newcommand{\be}{\begin{equation}}
\newcommand{\ee}{\end{equation}}
\newcommand{\bea}{\begin{eqnarray}}
\newcommand{\eea}{\end{eqnarray}}
\newcommand{\beann}{\begin{eqnarray*}}
\newcommand{\eeann}{\end{eqnarray*}}
\newcommand{\beasn}{\begin{sneqnarray}}
\newcommand{\eeasn}{\end{sneqnarray}}
\newcommand{\ba}{\begin{array}}
\newcommand{\ea}{\end{array}}
\newcommand{\nn}{\nonumber}
\newcommand{\Appendix}[1]%
    {%
     \section{#1}%
      }

\catcode`@=11
\def\secteqno{\@addtoreset{equation}{section}%
\def\theequation{\arabic{equation}}}
\def\endsecteqno{\def\theequation{\@ifundefined{chapter}%
{\arabic{equation}}{\thechapter.\arabic{equation}}}}
\newcounter{subequation}
\def\thesubequation{\alph{subequation}}
\def\sneqnarray{\stepcounter{equation}\let\@currentlabel=\theequation
\setcounter{subequation}{1}
\def\@eqnnum{{\rm (\theequation\thesubequation)}}
\global\@eqcnt\z@\tabskip\@centering\let\\=\@eqncr\let\@@eqncr=\@@sneqncr
$$\halign to \displaywidth\bgroup\@eqnsel\hskip\@centering
 $\displaystyle\tabskip\z@{##}$&\global\@eqcnt\@ne
 \hskip 2\arraycolsep \hfil${##}$\hfil
 &\global\@eqcnt\tw@ \hskip 2\arraycolsep
$\displaystyle\tabskip\z@{##}$\hfil
  \tabskip\@centering&\llap{##}\tabskip\z@\cr}
\def\endsneqnarray{\@@sneqncr\egroup $$\global\@ignoretrue}
\def\@@sneqncr{\let\@tempa\relax
   \ifcase\@eqcnt \def\@tempa{& & &}\or \def\@tempa{& &}
   \else \def\@tempa{&}\fi
     \@tempa \if@eqnsw\@eqnnum\stepcounter{subequation}\fi
     \global\@eqnswtrue\global\@eqcnt\z@\cr}
\def\nobiblabels{\def\@lbibitem[##1]##2{\@bibitem{##2}}}
\catcode`@=12

\def\a{\alpha}  \def\b{\beta} \def\g{\gamma} \def\G{\Gamma}
\def\d{\delta}  \def\e{\epsilon}

   \def\m{\mu} 
    
\newcommand{\PR}[3]{{\sl Phys. Rev.} {\bf #1} (19#2) {#3}}

\def\zb{\bar{z}}

      \def\ap3{a^+_3}
      \def\am3{a^-_3}
         \def\b3{a^3_3}

\def\dpz3{d^+_{3z}}   \def\dpzb3{d^+_{3\zb}}
\def\dmz3{d^-_{3z}}   \def\dmzb3{d^-_{3\zb}}
\def\ez3{d^3_{3z}}    \def\ezb3{d^3_{3\zb}}

\def\mm3{m^{-3}}
\def\mp3{m^{+3}}

\def\md3{m^{33}}

\def\bfnabla{\mbox{\boldmath $\nabla$}}
\def\bfsigma{\mbox{\boldmath $\sigma$}}

\documentstyle[12pt,epsf]{article}

\secteqno
\begin{document}


\title{{\bf The Lamb Shift in Dimensional Regularisation  }}
\author{{\Large {\sl Antonio Pineda}
                \ and \ {\sl Joan Soto}}\\
        \small{\it{Departament d'Estructura i Constituents
               de la Mat\`eria}}\\
        \small{\it{and Institut de F\'\i sica d'Altes Energies}}\\
        \small{\it{Universitat de Barcelona, Diagonal, 647}}\\
        \small{\it{E-08028 Barcelona, Catalonia, Spain.}}\\
        {\it e-mails:} \small{pineda@ecm.ub.es, soto@ecm.ub.es} }
\date{\today}

\maketitle
\thispagestyle{empty}


\begin{abstract}
We present a simple derivation of the Lamb shift
using effective field theory techniques and dimensional regularisation.
\end{abstract}

\bigskip
Keywords: Effective Field Theories, NRQED, NRQCD, HQET.

PACS: 11.10.St, 12.20.-m, 12.39.Hg

\vfill
\vbox{
\hfill{hep-ph/9711292 \null\par}
\hfill{UB-ECM-PF 97/33}\null\par}

\newpage



The explanation of the Lamb shift \cite{Lamb} is one of the early applications of the
quantised electromagnetic field \cite{Bette}. Although many text books \cite{Books}
contain detailed expositions on this effect, the conventional derivations are subtle
 and rather
intrincated. In particular a careful separation of small and large
momenta must be carried out at some stage of the calculation.

It has already been noticed \cite{Labelle1} that modern techniques of effective 
field theory lead to a simpler
derivation of the Lamb shift. The important point to be
realised is that in non-relativistic bound states there is a hierarchy of well
separated scales \cite{Lepage}. Namely, the mass of the particle which forms the 
bound state (hard),
the typical relative momentum in the bound state (soft) and the typical energy of the
bound state (ultrasoft). In ref. \cite{Labelle2} counting rules were given 
to work out the size of a given NRQED diagram in the Coulomb gauge for soft and
ultrasoft photons respectively. These rules were used in ref. \cite{Labelle1} 
to obtain
the Lamb shift $O(m\a^5)$, which requires the matching of QED to NRQED at one loop.
The calculations were done using a cut-off and a photon mass to regulate the UV 
divergencies in the
effective theory and
IR divergences in both theories respectively.  

In this note we point out that a further step in clarity and simplicity is achieved by
(i) matching NRQED to an effective theory for ultrasoft photons which we have called
potential NRQED (pNRQED) \cite{Mont} and (ii) using dimensional regularisation (DR) for 
both UV and IR
divergences.
Since we are interested in the physics at the ultrasoft scale, 
we may sequentially  integrate out  energies and momenta at the hard
and soft scales. This is carried out by matching to suitable effective field theories.
Integrating out the hard scale leads  from QED to the celebrated NRQED. Integrating
out the soft scale produces potential terms and leads from NRQED to 
pNRQED. In both cases the matching is most efficiently
done in DR. 
 The matching from QED to NRQED at one loop for the bilinear term in fermions 
has been carried out using dimensional regularisation in \cite{Manohar} and  
for the four-fermion operators 
in \cite{Mont}. Since we are interested in the interaction with a static source
 the infinite mass limit for the nucleus field is taken. Hence the 
only relevant terms at $O(m\a^5)$ in the NRQED Lagrangian are the following. 

\bea
{\cal L}_{NRQED}=&& \psi^\dagger \Biggl\{ i D^0 + \, {{\bf D}^2\over 2 m} +
 {{\bf D}^4\over
8 m^3} + c_F\, g {{\bf \bfsigma \cdot B} \over 2 m} 
 + c_D\, g {\left[{\bf \bfnabla \cdot E }\right] \over 8 m^2} \nonumber \\
&&
 +
 i c_S\, g {{\bf \bfsigma \cdot \left(D \times E -E \times D\right) }\over 8 m^2} 
\Biggr\} \psi  
+ N^{\dagger} iD^0 N \\
&& - {1\over 4}d_1 F_{\mu \nu} F^{\mu \nu}  
+{d_2\over m^2} F_{\mu \nu} D^2 F^{\mu \nu}\,, 
\nonumber
\eea
where $ i D^0=i\partial_0 -eA^0$ , $i{\bf D}=i{\bfnabla}+e{\bf A}$ on the electron
field $\psi$ and $ i D^0=i\partial_0 +eZA^0$ on the nucleus field. The matching
coefficients read
\begin{eqnarray} \label{26}
c_F &=& 1+{\alpha\over 2 \pi} \nn \,,\\
c_D &=& 1+{\alpha\over \pi} \left( 
{8\over 3} \log {m \over \mu} \right) \nn \,,\\
c_S &=& 1+{\alpha\over \pi} \,,\\ 
d_1 &=& 1-{ \alpha\over 3 \pi}  \log m^2/\mu^2 \nonumber \,,\\
d_2 &=& {\alpha\over 60 \pi}  \nn \,.
\end{eqnarray}
The fact that only the terms displayed in (1) are relevant can be easily seen by
 drawing diagrams of one electron one nucleus irreducible
Green function (i.e. diagrams which cannot be disconnected by cutting one electron
and one nucleus line) and taking into account that the next relevant scale is $m\a$. For
diagrams which cannot be disconnected by cutting a photon line the
order of the leading contribution of each diagram is $m\a^{r+s+1}$, where $r$ is the
number of $\a$ that appear explicitly in the diagram and $s$ the number of $1/m$
factors (which must be compensated by $m\a$ until we obtain dimensions of energy). For
diagrams which can be disconnected by cutting a photon line there is an extra
suppression if $n$ time derivatives act on this photon line. This is due to the fact
that these time derivatives are only sensible to the typical energy. The
extra suppression
factor is $\a^{n}$.     
We should keep in mind however that
 a given diagram in NRQED 
 in general contains subleading contributions in $\a$
as well \cite{Labelle2}. On the contrary pNRQED will be specially designed in such a
 way that each term
and diagram contributes to a single order in $\a$.

Before going on, let us briefly comment on $d_1$ in (1). This factor can be set to one by 
rescaling
the photon field. The only effect this rescaling  has in the effective Lagrangian
 is redefining the electric charge $e$. The
redefined $e$ is the low energy electron charge which is the one actually measured in
low energy experiments, namely the one whose value is
$e^2/4\pi=\a\sim 1/137$. Therefore $d_1$ will be put to one from now on.

 pNRQED must be regarded as an effective field theory where electron and photon
 energies and
(relative) momentum $\sim m\a$ have been integrated out.
It is important to realise that because both the energy and momentum we are integrating
 out are $\sim m\a$ we can use
HQET (static) propagators for the electron field  to carry out the matching.
 This allows to match NRQED and pNRQED at a given order of
$1/m$ and $\a$.
The pNRQED Lagrangian can be written down at two levels: (i) in terms of electron and
nucleus fields and (ii) in terms of bound state wave function fields. The level (i)
is suitable for the matching calculation whereas the level (ii) is more convenient
for bound state calculations. In fact it is at this level where the order of
each term becomes explicit.  
The pNRQED (i) Lagrangian 
is a functional of $ \psi (t,{\bf x}) $, $N(t,{\bf x})$ and
ultrasoft photon fields $ A_{\m} (t,{\bf x}) $, which also contains (potential) terms
non-local in space. $ A_{\m} (t,{\bf x}) $ being ultrasoft  
means that they must be
 multipole
expanded about  
the position of the nucleus
in the one electron one nucleon Green function.
Recall that this is due to the fact that
 in a bound state ${\bf x} \sim 1/m\a$ whereas the
 typical 
energy and momentum scales for ultrasoft photons are $\sim m\a^2$.

The matching for the electron and nucleus bilinears is trivial 
since the related Green functions are blind to the
relative momentum. Hence the terms bilinear in electron fields and the terms bilinear in
nucleus fields in NRQED remain the same in pNRQED.
The
terms containing photon fields only also remain exactly the same in NRQED and pNRQED.
 However 
we should keep in mind that now they correspond to ultrasoft photons only.  
The matching of one electron one nucleus Green functions induces potential
 terms in
pNRQED. Namely terms bilinear in the electron and nucleus fields which are local in
time but non-local in space. If we use the Coulomb gauge, at the order we are
interested in it is enough to calculate the above Green functions at zero 
energy  and at some non-zero relative momentum ${\bf p} $. Indeed, if the photon
fields in this Green function
calculated from the pNRQED 
Lagrangian are
multipole expanded, any loop gives zero in DR as we argue next. First of all
there is no scale for the energy. Second, there is a scale for the momentum but
 both the HQET propagators and the (multipole expanded) transverse gluon propagators 
are insensitive to it. Hence all loops in pNRQED have no scale and can be put to zero 
in DR. Consequently the potential terms in pNRQED can be read off directly from the
calculation in NRQED.

A word of caution is needed. Strictly speaking the procedure above is not complete:
 it misses off-shell photons of
energy $\sim$ $m\a^2$ and momentum $\sim$ $m\a$. In particular, we miss the 
contribution of diagrams which can be disconnected by cutting a photon line 
when time derivatives appear in this line. 
In the Coulomb gauge such photons start playing a role at order $m\a^7$.
However, if a covariant gauge is used longitudinal photons in that region start
playing a role already at order $m\a^4$ and hence they must be taken into account to
calculate the Lamb shift in those gauges. Anyway, the matching procedure can be refined
so that the effect of such photons is included. We shall elaborate on this point
elsewhere.

At the order we are interested in only the NRQED diagrams in Fig. 1 
contribute to 
the matching. We obtain

\bea
L_{pNRQED}=&& \int d^3{\bf x}\Biggl(\psi^{\dagger} \Biggl\{ i D^0 + \,
 {{\bf D}^2\over 2 m} +
 {{\bf D}^4\over
8 m^3} 
\Biggr\} \psi 
 + N^{\dagger} iD^0 N 
  - {1\over 4} F_{\mu \nu} F^{\mu \nu} \Biggr) \\
&&+\int d^3{\bf x}_1 d^3{\bf x}_2  N^{\dagger} N (t,{\bf x}_2 ) \Biggl(
 {Z \a \over \vert {\bf x}_1-{\bf x}_2\vert} +
{Ze^2\over m^2}\left(-{c_D\over 8} +4d_2\right)\d^3 ({\bf x}_1-{\bf x}_2 ) \nn\\
&& +ic_S{Z\a\over 4m^2}{\bfsigma} \cdot \left({{\bf x}_1-{\bf x}_2 \over \vert {\bf x}_1-{\bf x}_2 
\vert^3}\times {\bfnabla} \right)\Biggr)\psi^{\dagger}\psi (t,{\bf x}_1 )
\nn \,.\eea

It is convenient to project the Lagrangian (at level (i)) above to the one electron 
one nucleus
subspace and thus obtain the pNRQED Lagrangian at level (ii).
 This subspace is spanned by
\be
 \int d^3{\bf x}_1 d^3{\bf x}_2 \varphi ({\bf x}_1,{\bf x}_2 ) \psi^{\dagger}({\bf x}_1)
 N^{\dagger}({\bf x}_2) \vert 0 \rangle
\,, 
\ee
where $\vert 0 \rangle$ is the subspace of the Fock space containing zero electrons and
nuclei but an arbitrary number of ultrasoft photons. Since there are not 
spatial derivatives acting on ${\bf x}_2$ the nucleus can be approximated 
by a
static source which may be taken at the origin. Then
\be
\varphi ({\bf x}_1,{\bf x}_2 )=\varphi ({\bf x}_1)\delta^3({\bf x}_2 )
\,.
\ee

The dynamics of $\varphi (t,{\bf x})$ is then described by the following Lagrangian
\bea 
&&L_{pNRQED}= \int d^3{\bf x} \varphi^{\dagger} (t,{\bf x})\Biggl(i\partial_0-e(1-Z)
A_{0} (t,{\bf 0})-ex^i\partial_i A_{0} (t,{\bf 0})
+ {{{\bfnabla}}^2\over 2 m}+{Z \a \over \vert {\bf x}\vert}\nn\\ 
\nn &&
 -ie{{\bf A} (t,{\bf 0}) \cdot {{\bfnabla}}\over  m}
 +{{{\bfnabla}}^4\over
8 m^3} 
+{Ze^2\over m^2}\left( -{c_D\over 8} +4d_2\right)\d^3 ({\bf x}) 
 +ic_S{Z\a\over 4m^2}{\bfsigma} \cdot \left({{\bf x}\over \vert {\bf x} 
\vert^3}\times {\bfnabla} \right)
\Biggr)\varphi (t,{\bf x}) 
\\ && 
- \int d^3{\bf x} {1\over 4} F_{\mu \nu} F^{\mu \nu}
\,,
\eea
where we have multipole expanded $ A_{\m} (t,{\bf x}_1) $ about zero.
Each term in this Lagrangian has a well defined size. $\partial_{i}$ acting on
the electron field and $1/|{\bf x}|$ are $\sim m\a$. Space derivatives acting 
on $ A_{\m} (t,{\bf 0}) $,
time derivatives and $ A_{\m} (t,{\bf 0}) $ itself are $\sim m\a^2$.
  The contribution of a given diagram to the energy is given by
$m\a^{r+2s+p+2}$. $r$ is the number of explicit $\a$ in the diagram, $s$ the number of
$1/m$ and $p$ the number of $x^{i}$ minus three times the number of $\d^3({\bf x})$
minus the number of $1/\vert {\bf x}\vert$ minus the number of $\partial_{i}$. We have
kept in (6) only the terms which are relevant to calculate the energy at $O(m\a^5)$.
There are more terms of the same size (e.g. $-ie({\bf x .\bfnabla}{\bf A} (t,{\bf 0}))
{\bf . \bfnabla} /  m $) which however do not contribute to the energy at the
 desired order
as it can be seen by counting the order of the diagrams in which they appear. 
(Dropping the
term $-ex^ix^j\partial_i\partial_j A_{0} (t,{\bf 0})$ requires a more detailed analysis
 if a covariant gauge is to be used).

 Gauge invariance 
can be
checked order by order in this counting
by introducing a gauge covariant field $S(t,{\bf x})$ as $\varphi (t,{\bf x})=
 \bigl[e^{ie\int_{{\bf 0}}^{{\bf x}} {\bf A} d{\bf x}}
\bigr]S (t,{\bf x} )
 $, such that it transforms
$S(t,{\bf x}) \rightarrow e^{-ie(Z-1)\theta (t,{\bf 0})}S(t,{\bf x})$. From this
transformation it is clear that we should regard $S(t,{\bf x})$ like an ion field
rather than an electron field. After multipole expanding
at the desired order, we obtain an explicitly gauge invariant Lagrangian

\bea
&L_{pNRQED}=&\int d^3{\bf x} S^{\dagger} (t,{\bf x})\Biggl(i\partial_0-e(1-Z)
A_{0} (t,{\bf 0})+e{\bf x} \cdot {\bf E}  (t,{\bf 0})
+ {{{\bfnabla}}^2\over 2 m} 
+{Z\a\over \vert {\bf x}\vert}
 \nn
\\&&
+ {{{\bfnabla}}^4\over
8 m^3}
 +{Ze^2\over m^2}\left( -{c_D\over 8} +4d_2\right)\d^3 ({\bf x}) 
 +ic_S{Z\a\over 4m^2}{\bfsigma} \cdot \left({{\bf x}\over \vert {\bf x} 
\vert^3}\times {\bfnabla} \right)
\Biggr)S (t,{\bf x})
\nn
\\ && 
- \int d^3{\bf x} {1\over 4} F_{\mu \nu} F^{\mu \nu}
\,.
\eea

It is also interesting to notice that the same Lagrangian above can be obtained by
using backwards the equation of motion in the first term of the second line in (6).

\bea
&&\int d^4x \varphi^{\dagger}i{e\over m}{\bf A} (t,{\bf 0}) \cdot 
{\bfnabla}\varphi= 
\int d^4x \varphi^{\dagger}ie{\bf A} (t,{\bf 0}) \cdot \left[{\bf x},\hat h_0\right]\varphi= 
\int d^4x \varphi^{\dagger}e \partial_0 {\bf A}  (t,{\bf 0}) \cdot {\bf x} \, \varphi
\nn \,,\\
&&
\eea
where 
\be
\hat h_0=-{\bfnabla^2\over 2m}-{Z\a\over\vert {\bf x} \vert }
\,.
\ee

The Lamb shift receives contributions from the propagation of ultrasoft photons 
in the bound state. Notice that since the lagrangian of pNRQED (ii) is explicitly gauge
invariant we can use any gauge to calculate with it. Still the Coulomb gauge continues
to be advantageous. In this gauge $A_0(t,{\bf 0})$ can be dropped if we use DR since
it only produces tadpoles. If a covariant gauge is used the propagation of
 $A_0(t,{\bf 0}
)$ must also be taken into account. In the Coulomb gauge
at $O(m\a^5)$ the ultrasoft photons contribute to the bound state energy 
only through the pNRQED diagram of Fig. 2. In order to calculate this 
contribution we consider the two point function

\be
\Pi(q,{\bf x}):=\int d x^0 e^{iqx^0}<T\{ \varphi (0) \varphi^{\dagger}(x) \}>
\ee  
when $q \rightarrow E_{n}$. We write

\be
\Pi(q,{\bf x})={A_n +\d A_n \over q-(E_n+\d E_n )}\sim {A_n +\d A_n \over q-E_n}+
{A_n \over q-E_n}\d E_n{1 \over q-E_{n}}
\,.
\ee

$\Pi(q,{\bf x})$ must be calculated in $D$ dimensions. By introducing the complete set of
eigenfunction of $\hat h_0$ in $D=4-\epsilon$ dimensions $\hat h_0 \phi_{n}= 
E_{n}\phi_{n}$, we
obtain

\bea
\Pi (q,{\bf x}) \sim {i\phi_{n}(0)\phi_{n}^{\dagger}({\bf x})\over  q-E_{n}} 
-{e^2
}
{i\phi_{n}(0)\over  q-E_{n}}\sum_{m}
\langle n\vert {\bf v}^i \vert m\rangle 
I_{ij}(q-E_m)
\langle m\vert {\bf v}^j \vert n\rangle
{i\phi^{\dagger}_{n}({\bf x})\over  q-E_{n}}
\,,
\eea
where ${\bf v}=-i{\bfnabla}/m$. We have then to calculate the following integral

\bea
I_{ij}(p)=&&\int {d^D k\over (2\pi)^D} {i\over k^2}\left( \d^{ij}-{k^i k^j
\over {{\bf k}}^2}\right){i\over p-k^0+i\eta }\nn\\
&&\\  = &&
-ip{1\over 6\pi^2}\d^{ij}\left({1\over \e}+{1\over 2}\log{ 4\pi}+\log{
\m\over
-p-i\eta}+{5\over 6}-{\g\over 2}-\log{2}
\right)\nn
\,.
\eea  
This integral is carried out by (i) first integrating over $k_0$ and then (ii) 
choosing a contour
in the $\vert{\bf k}\vert$ complex plain which embraces the cut on the positive 
real axes and closes itself at infinity in such a way that the pole of the integrand
is always left inside the contour.
Notice that the divergent part of this integral is a polynomial. Before going on let us
check that it can be absorbed by a counterterm in $L_{pNRQED}$. Indeed,
\bea
\d^{US,local} E_{n} \sim
 && \sum_{m}
\langle n\vert {\bf v} \vert m\rangle 
\left( q-E_{m}\right)
\langle m\vert {\bf v} \vert n\rangle 
=  
\langle n\vert {\bf v}  
\left( q-\hat h_0 \right)
 {\bf v} \vert n\rangle 
  \\
= &&
\langle n\vert 
\Big( 
{1\over 2}\left( q-\hat h_0 \right) 
{\bf v}^2+{1\over 2}{\bf v}^2\left(q-\hat h_0 \right)
-{1\over 2}\left[
{\bf v}\left[{\bf v} ,q-\hat h_0 \right]\right] \Big)
\vert n\rangle
 \nn
\,.
\eea
In order to identify $\d E_{n}$ the strict limit
$q\rightarrow E_{n}$ must be taken and hence the first two terms in the last expression
can be dropped. The remaining term is nothing but the $D-1$ dimensional Laplacian
acting on the $D$ dimensional Coulomb potential which gives a $D-1$ dimensional
$\d$-function. Thus (14) reads
\be
\d^{US,local} E_{n} \sim 
\langle n \vert 
\left( -{Ze^2\over 2m^2} \d^{D-1}({\bf x } )  \right)
\vert n \rangle 
\,,
\ee
 which can be absorbed by a counterterm in the
second last term of $L_{pNRQED}$ in (6) and (7). Since we have used $\overline{MS}$
 subtraction
scheme in the matching we must use the same subtraction scheme here.
 Since there are no further
divergences left the limit $D\rightarrow 4$ can be safely taken. Notice that this
procedure avoids calculating explicitly $\phi_{n}$ and $E_{n}$ in $D$ dimensions.
We finally obtain that the contribution to the energy given by the diagram in Fig. 2 is

\bea
\d^{US} E_{n} = && {2\over 3}{\a\over \pi}\Bigg(\left( \log{\m\over m}+{5\over
6}-\log{2}\right)\left({Ze^2\over 2}\right){\vert \phi_{n}({\bf 0})\vert^2 \over m^2}
\nn\\
 && \\
&&
-\sum_{m\not= n}
\vert
\langle n\vert {\bf v}\vert m \rangle
\vert^2
\left( E_{n} -E_{m}\right)\log{m\over\vert  E_{n} -E_{m}\vert} \Bigg)
\nn
\,.
\eea

From the imaginary part of (13) we also obtain the total width
\be
\G_n=\sum_{m<n}{4\over 3}\a
\vert
\langle n\vert {\bf v}\vert m \rangle 
\vert^2 \left(E_n -E_m \right)
\,.
\ee

 The remaining
contribution at $O(m\a^5)$ arises from the above mentioned potential terms in (6)
and (7).
 It reads
\bea
\d^{S} E_{n} =&&\d^{S,K} E_{n} +\d^{S,\d} E_{n} +\d^{S,S} E_{n}\nn \,,\\&&\\
\d^{S,K} E_{n} =&&-{1 \over 8m^3} \langle nlj\vert \bfnabla^4 \vert nlj \rangle \nn \,,\\
\d^{S,\d} E_{n} =&&{Ze^2\over m^2} \left( {c_D\over 8}-4d_2 \right) \vert
\phi_{n}({\bf 0})\vert^2 \nn \,,\\
 \d^{S,S} E_{n}  =&&c_S{Z\a\over 4m^2} \left( j(j+1)-l(l+1)-{3\over 4}\right)
\langle nlj\vert {1\over{\bf x}^3}\vert nlj \rangle 
\nn 
\eea
($\vert n \rangle =\vert nlj \rangle $).
Thus the correction to the bound state energy $\d E_n$ up to $m\a^5$ included is
obtained by adding up the soft and ultrasoft contributions
\be
\d E_n =\d^{S} E_{n} + \d^{US} E_{n}
\,,
\ee
which agrees with the well known result. Recall that the hard contributions
are encoded in $c_D$, $c_S$ and $d_2$.
Notice also that the subtraction point dependences
of $\d^{S,\d} E_{n}$ (in $c_D$) and $\d^{US} E_{n}$
cancel out, as they should. However we should stress that in order to obtain the 
finite pieces right, it is
 necessary that the same subtraction scheme, namely $\overline {MS}$, which has been
used in the matching be used in the calculations of pNRQED.

In summary we have presented a simple and complete derivation of the Lamb shift
based on modern EFT
techniques and dimensional regularisation. The derivation goes through two EFTs,
namely NRQED and pNRQED, so that at any stage it becomes clear which
terms and diagrams are to be considered to carry out a calculation at the desired
 order. The formulation of pNRQED, namely an effective field theory for ultrasof
photons, has received quite some attention lately \cite{Labelle2,Grinstein}. 
In ref. \cite{Mont} the matching from NRQED (NRQCD) to pNRQED (pNRQCD) was 
outlined and the pNRQED and pNRQCD Lagrangians for positronium and quarkonium 
were presented. Here we have worked out the pNRQED Lagrangian for Hydrogen-like 
atoms and presented its first application. 
 Gauge invariance is manifest in the two EFTs. This is particularly remarkable in
pNRQED, if we take into account that most of the previous attempts to deal with ultrasoft 
photons were
strongly based on the Coulomb gauge. Last but not least DR allows to regulate both UV
 and IR divergences
in a manifestly gauge invariant way (unlike the photon mass) and makes the matching
calculations straightforward.
 These techniques are most promising when applied to bound states of heavy
quarkonia in QCD \cite{Mont}.

\section*{Acknowledgements}

A.P. acknowledges a grant from the generalitat de Catalunya. Financial
support from CICYT, contract AEN95-0590 and from CIRIT, contract
GRQ93-1047 is also acknowledged.
   


\vfill
\eject

{\bf Caption1.} The first diagram is the Coulomb potential. The circle is the 
vertex proportional to $c_D$, the square to 
$c_S$ (spin dependent) and the dashed dot to $d_1$ (the vacuum polarization). 
The dashed, solid and double lines are the static photon, electron and nucleus 
propagators respectively.
\bigskip

{\bf Caption2.} The thick line and wavy line are the ion and transverse 
photon propagators respectively.

\vfill
\eject

\medskip
\begin{figure}
\hspace{1.5in}
\epsfxsize=5.0in
\centerline{\epsffile{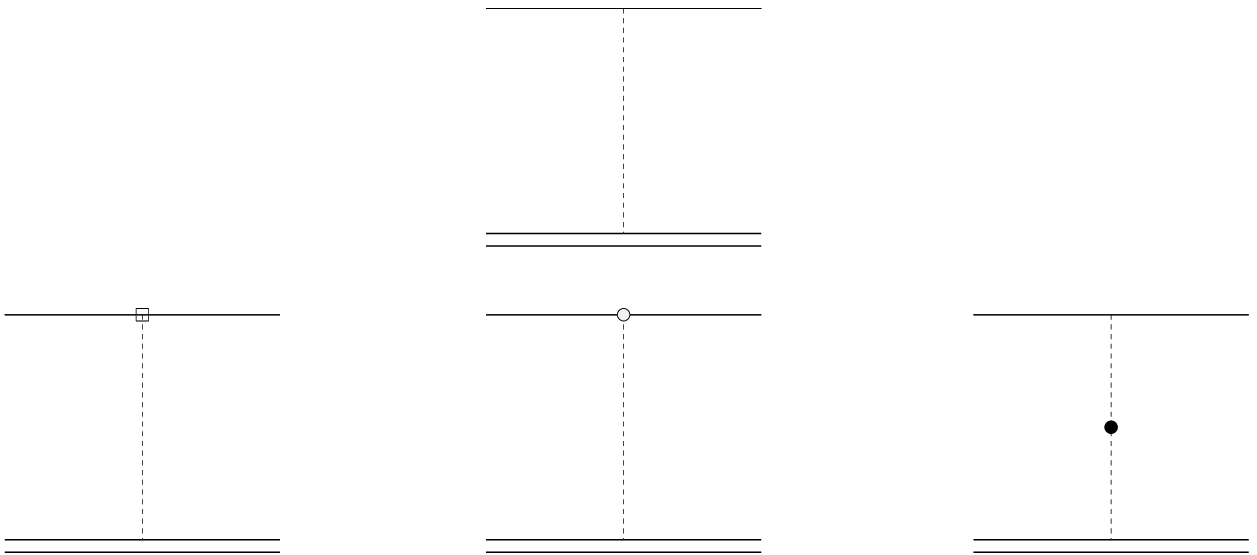}}
\caption{}
\label{fig1}
\end{figure}
\medskip

\medskip
\begin{figure}
\hspace{0.7in}
\epsfxsize=2.5in
\centerline{\epsffile{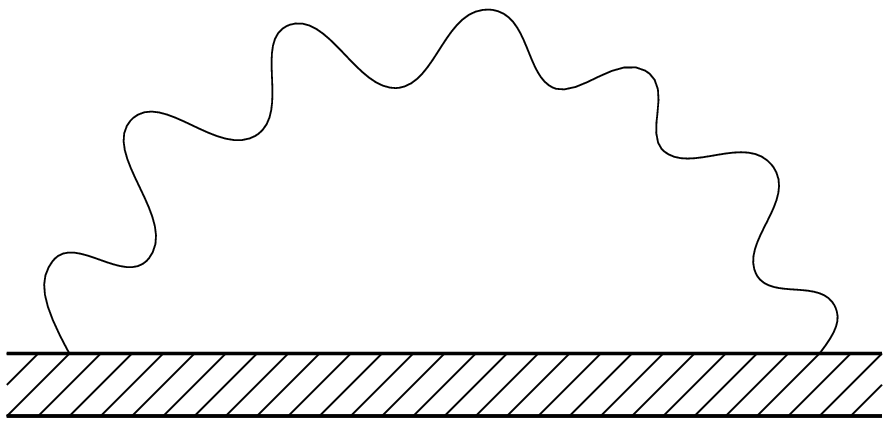}}
\caption{}
\label{fig2}
\end{figure}
\medskip


\begin{thebibliography}{99}

\bibitem{Lamb} W.E. Lamb and R.C. Retherford, \PR{72}{47}{241}.

\bibitem{Bette} H.A. Bethe, \PR{72}{47}{339}. 

\bibitem{Books} C. Itzykson and J.-B. Zuber, {\it Quantum Field Theory} 
(McGraw-Hill, 1980).

\bibitem{Labelle1} P. Labelle and S.M. Zebarjad, {\it Derivation of the
Lamb Shift Using an Effective Field Theory}, McGill-96/41, 
hep-ph/9611313.

\bibitem{Lepage} W.E. Caswell and G.P. Lepage, {\it Phys. Lett.} {\bf B167}
(1986) 437.

\bibitem{Labelle2} P. Labelle, {\it Effective Field Theories for QED Bound
States: Extending Nonrelativistic QED to Study Retardation
Effects}, McGill-96/33, hep-ph/9608491.

\bibitem{Mont} A. Pineda and J. Soto, {\it Effective field Theory for ultrasoft
momenta in NRQCD and NRQED}, to be published in the proceedings of QCD97: 
25th Anniversary of QCD, Montpellier, France, 3-9 July 1997, UB-ECM-PF 97/17, 
hep-ph/9707481.

\bibitem{Manohar}  A.V. Manohar, {\it Phys. Rev.} {\bf D56} (1997) 230. 

\bibitem{Grinstein} I.Z. Grinstein and I.Z. Rothstein, {\it
Phys. Rev.} {\bf D57} (1998) 78. 
M. Luke and A.V. Manohar, {\it Phys. Rev.} {\bf D55} (1997) 4129. M. Luke and M.J. Savage, 
{\it Power counting in dimensionally regulated NRQCD}, UTPT-97-12, 
hep-ph/9707313.

\end{thebibliography}
\end{document}